# Discovery of a high confident soft lag from an X-ray flare of Markarian 421

ZHANG YouHong[1†], HOU ZhengTian[2], SHAO Li[3] & ZHANG Peng[1]

1. Department of Physics and Tsinghua Center for Astrophysics (THCA), Tsinghua University, Beijing 100084, China;
2. Group of Mathematics and Physics, Shangqiu Medical College, Beihai Road No. 486, Shangqiu 476100, Henan Province, China;
3. Max-Planck-Institut für extraterrestrische Physik, Postfach 1312, Garching 85741, Germany

**We present the X-ray variability properties of the X-ray and TeV bright blazar Mrk 421 with a ~60 ks long XMM-Newton observation performed on 2005 November 9–10. The source experienced a pronounced flare, of which the inter-band time lags were determined with very high confidence level. The soft (0.6–0.8 keV) X-ray variations lagged the hard (4–10 keV) ones by $1.09^{+0.11}_{-0.12}$ ks, and the soft lag increases with increasing difference of the photon energy compared. The energy-dependent soft lags can be well fitted with the difference of the energy-dependent cooling timescales of the relativistic electron distribution responsible for the observed X-ray emission, which constrains the magnetic field strength and Doppler factor of the emitting region as $B\delta^{1/3} \sim 1.78$ Gauss.**

BL Lacertae objects: individual: Mrk~421, galaxies: active, radiation mechanisms: non-thermal, X-rays: galaxies

The X-ray and TeV emission from high-energy peaked BL Lac objects (HBLs) is thought to be radiated by the most energetic end of the same electron population via synchrotron and synchrotron Self-Compton (SSC) processes, respectively. The X-ray and TeV variations of HBLs is therefore expected to be highly correlated and to be most variable across the whole electromagnetic wavelengths. One of the most important goals of studying HBLs emitting TeV photons (so-called TeV blazars) is to explore the accelerating and cooling mechanisms of the high energy electrons responsible for the observed X-ray and TeV emission. A qualitative but meaningful way to achieve such goal is to measure time lags of the variations between different X-ray and TeV energies and between X-ray and TeV bands, which provides useful information about the acceleration and cooling timescales of high energy electrons and constrains physical parameters of the emitting region. The ratio of acceleration to cooling timescales of the high energy electrons determines the sign of lag between the variations of different energies, which is observed as soft (the lower energy variations lag the higher energy ones) or hard (the higher energy variations lag the lower energy ones) lag [1–2]. In fact, both signs of lag have been claimed for a few bright HBLs either in X-ray (e.g., PKS 2155-304[2–4]; Mrk 421[5–7]) or in TeV (Mrk 501[8]) energies. However, the reliability of the lags derived from the data obtained with low Earth orbit satellites was questioned by Edelson et al. [9] because the reported lags are usually smaller than the orbital periods of the satellites. This argument was later disproved by Zhang et al. [10], who demonstrated that periodic data gaps indeed introduce larger uncertainties on lag measurements than evenly sampled data, but the lags of sub-hour can not be the artificial results of periodic gaps. The discoveries of lags on the order of sub-hour with evenly sampled XMM-Newton data of Mrk 421[7] and PKS 2155-304[3] supported the realities of the lags previously found by low Earth orbit satellites ASCA and BeppoSAX.

Mrk 421 is the brightest and best studied HBL in both X-ray and TeV bands. As a calibration target, XMM-Newton repeatedly monitored the source since 2000. This paper presents a detailed temporal analysis of the PN observation performed on 2005 November 9–10, to show the significant determination of inter-band soft lags of the X-ray variations from a pronounced X-ray flare obtained during this observation.

Received; accepted
doi:
†Corresponding author (email: youhong.zhang@mail.tsinghua.edu.cn)
Supported by the National Basic Research Program of China − 973 Program 2009CB824800 and the National Natural Science Foundation of China (Project 10878011, 10733010 and 10473006) and the Key Project of Chinese Ministry of Education (NO 106009) and Tsinghua University Project for top scholars

## 1 XMM-Newton Observation and data reduction

The observation was performed during XMM-Newton flight of orbit 1084 (ObsID = 0158971301) and lasted ~60 ks. The pn camera onboard XMM-Newton was operated in timing mode. The data were reprocessed with the XMM-Newton Science Analysis System (SAS) version 7.1.2 and with the latest calibration files as of 2008 May 11. The background is quite low and stable, and especially the PN data were not affected by the photon pile-up effect. We extracted PN source photons from rows $28 \leq RAWX \leq 44$, centered on the brightest strip of the source. The background photons were selected from rows $2 \leq RAWX \leq 11$ and $57 \leq RAWX \leq 63$. In case of PN timing mode, the ratio of single to double events depends on the source position, so we selected photons from the combined single and double events (i.e., PATTERN$\leq$4) with quality FLAG=0. The energy range is restricted to $0.6 \leq E \leq 10$ keV to avoid the increased noise below 0.6 keV. Very high count rate of Mrk 421 made PN detector frequently fall into counting mode. This introduced ``quasi-periodic'' gaps of typically ~21.6 s duration into the PN data stream. Totally, there are 1292 of these ``elementary observation intervals'' (EOIs)[11]. On average, each EOI has data of ~23.8 s long followed by a gap of ~21.6 s long, which, however, depends on the count rate due to strong flux variations of the source.

## 2 Results

The upper panel of Figure 1 plots the 600s binned 0.6–10keV band light curve. It can be seen that the source was highly variable, exhibiting a ~45% variability amplitude from the minimum to the maximum. From the beginning to ~16,000 s of the exposure, the light curve is characterized by a shallow decay with two ``flickers'' superimposed on it. From then on, the source experienced a pronounced flare, characterized by a quasi-symmetric profile if one does not consider the flickers superimposed on the rise and decay phase of the flare. The timescale of the rise phase is ~26,000 s, with a ``flicker'' superimposed on it. The decay phase may not be fully sampled due to stop of the exposure. A "flare" of smaller amplitude appeared after ~ 6,000 s of the main flare peak, which blurs the decay phase of the main flare. The X-ray variability of Mrk 421 is therefore quite complex: a main flare is usually contaminated by many small amplitude flickers. Also shown in Figure 1 (bottom panel) are the 600 binned 0.6–0.8 keV (soft) and the 4–10 keV (hard) band PN light curves, respectively. For direct comparisons, the soft and hard band light curves are normalized to their respective mean value.

The soft and hard band light curves follow each other very well. However, the hard light curve is more variable than the soft one. The increases of X-ray variability with increasing energy is already known in Mrk 421[7,12]. Similar behavior of variability is also frequently observed in other HBLs, e.g., PKS 2155-304[13]. Moreover, the main flare, defined as the time interval between the two dashed lines of Figure 1, shows some interesting features. As seen from the bottom panel of Figure 1, the 4–10 keV light curve clearly peaks before the 0.6–0.8 keV one does.

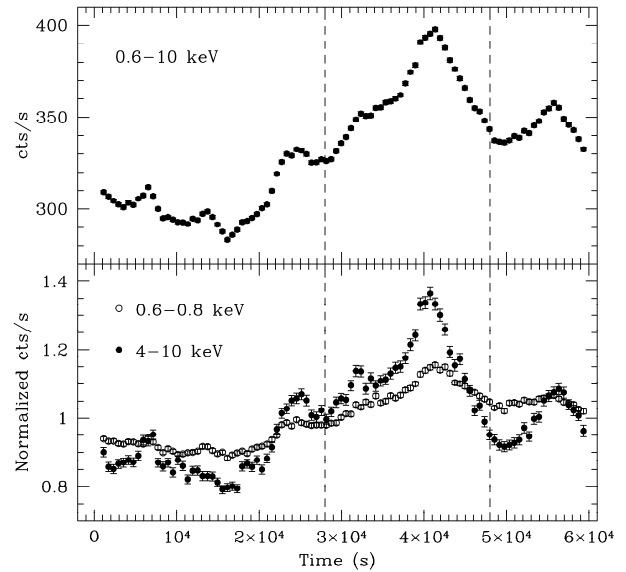

Figure 1. Top panel: the 600s binned 0.6–10 keV PN light curve. Bottom panel: the 600s binned 0.6–0.8 keV and 4–10 keV PN light curves. For direct comparison, the data are normalized to their respective mean. The time interval between the two dashed lines is defined as the main flare used for lag determinations in the text.

Thanks to long orbital period, data gathered with XMM-Newton can be continuous over exposure time of more than one day. This is very important to study intra-day X-ray variability of HBLs, like Mrk 421. The light curves obtained with PN timing mode will be evenly sampled if the data are binned over time longer than the EOI time span (~45.4 s). This binning ensures that each data bin is, on average, ~50% exposed. Time lags can be estimated by calculating standard Cross-Correlation Function (CCF) between two time series. We computed CCFs between light curves of different energy bands for the main flare only. The light curves are re-binned over 100 s. Figure 2 shows the CCF between the 0.6–0.8 keV and 4–10 keV light curves for the main flare. A positive lag indicates that the lower energy light curve lags the higher energy one. The CCF clearly peaks around 1000s, and the CCF shape is roughly symmetric with respective to the CCF peak. This indicates that the 0.6–0.8 keV variations lag the 4–10 keV ones by ~1000 s.

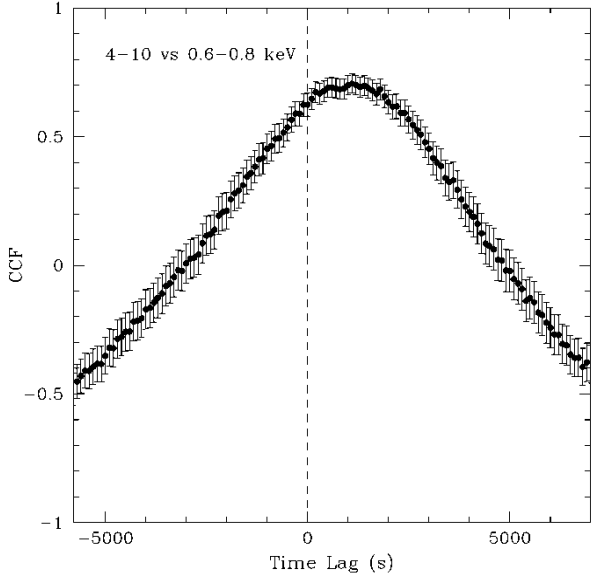

Figure 2. The CCF between the 4—10 keV and the 0.6—0.8 keV PN light curves for the main flare. Note that the lag range shown are [–6000, 7000] s only in order to clearly show quasi-symmetry of the CCF with respective to the positive lags of ~1000 s (the first curve lead the second one). The data are binned over 100s.

We apply Monte Carlo simulation to estimate the lag and its uncertainty[3], which is the median of the simulations and the corresponding 68% (1σ) confidence level errors. The lag is measured with three techniques: (1) $\tau_{peak}=1.10^{+0.30}_{-0.50}$ ks, is the lag corresponding to the CCF peak ($r_{max}$); (2) $\tau_{cent}=1.00^{+0.10}_{-0.11}$ ks, is obtained by computing the CCF centroid; (3) $\tau_{fit}=1.09^{+0.11}_{-0.12}$ ks, is the lag corresponding to the peak of an asymmetric Gaussian function (plus a constant) fitted to the CCF. After finding $r_{max}$ of the CCF, the corresponding $\tau_{peak}$ is recorded. Both $\tau_{cent}$ and $\tau_{fit}$ are then measured only using the CCF points with r in excess of $0.5r_{max}$. As the CCFs are roughly symmetric, $\tau_{peak}$, $\tau_{cent}$ and $\tau_{fit}$ are in good agreement with each other within the 1σ uncertainties. We repeat the same procedure to derive the lags of the 0.8–1, 1–2 and 2–4 keV light curves with respective to the 4–10 keV one, respectively. Figure 3 presents the energy dependence of soft lags measured with $\tau_{fit}$, clearly showing that the lag increases with increasing energy difference compared.

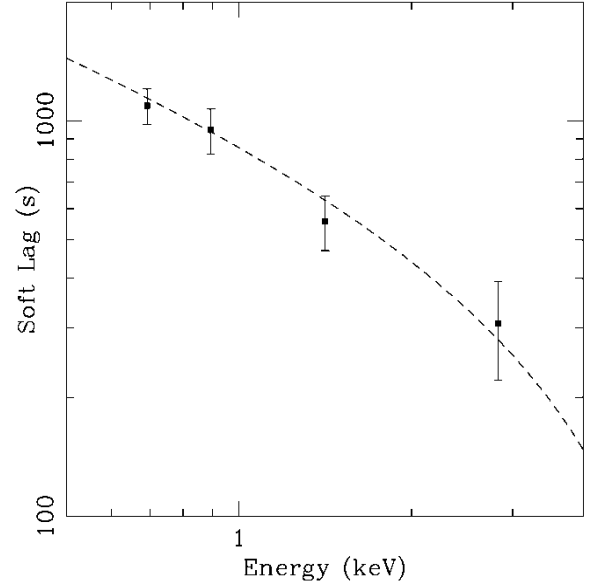

Figure 3. The observed energy-dependent soft lags, measured with $\tau_{fit}$ method, for the main flare. All lags are measured with respective to the 4–10 keV light curve. The dashed line is the best fit with Equation (4).

## 3 Discussion

Previous X-ray spectral analyses showed that the synchrotron emission of Mrk 421 peaked at ~1 keV[14,15]. Our preliminary spectral fit with a broken power law revealed convex X-ray spectral shape, with the soft and hard photon index of ~2.3 and ~2.6, respectively, indicating the synchrotron signature of the 0.6—10 keV emission for the observation presented in this paper.

In the commonly accepted shock-in-jet model[2,11], the blazar flare is thought to be produced by two-colliding-shells, from which the particles are accelerated and magnetic field is generated. For an accelerated electron population tangled with random magnetic field, the typical synchrotron emission frequency of an electron with energy $\gamma mc^2$ is $\nu \sim 3.73 \times 10^6 B\gamma^2$ Hz (averaged over pitch angles, γ is the Lorentz factor of the electron). Under the pure synchrotron emission scenario, the accelerating timescale ($t_{acc}$, assuming the diffusive shock acceleration) and cooling timescale ($t_{cool}$) of an electron can be approximately expressed in terms of the typical synchrotron photon energy of that electron[2,5]. In the observer's frame, we obtain

$$t_{acc}(E) = 9.65 \times 10^{-2}(1+z)^{3/2}\xi B^{-3/2}\delta^{-3/2}E^{1/2} \text{ s} \quad (1)$$

$$t_{cool}(E) = 3.04 \times 10^{3}(1+z)^{1/2} B^{-3/2}\delta^{-1/2}E^{-1/2} \text{ s} \quad (2)$$

where z is the red shift of a source, E (in keV) the observed photon energy, B (in Gauss) and δ are the magnetic field strength and Doppler factor of the emitting blob, respectively, and ξ describes how fast an electron is being accelerated.

Equations (1) and (2) show that $t_{acc}(E)$ and $t_{cool}(E)$ of electrons is photon energy-dependent in an opposite sense: higher energy electrons cool faster but accelerate slower than lower energy electrons do. If $t_{acc}(E) < t_{cool}(E)$, the emission propagates from higher to lower energy, the lower energy variations lag the higher energy ones (i.e., the so-called soft lag); In case of $t_{acc}(E) \sim t_{cool}(E)$, the emission propagates from lower to higher energy, so the higher energy variations lag the lower energy ones (i.e., so-called hard lag). As $t_{acc}(E) < t_{cool}(E)$ is always true for low energy electrons, a hard lag is observable only at energies close to the maximum synchrotron photon energies emitted by the maximum energy ($\gamma_{max}$) electrons under the particular scenario of $t_{acc}(\gamma_{max}) \sim t_{cool}(\gamma_{max})$[1]. As a result, soft lag is more frequently observable than hard lag[3]. For Mrk 421, hard lag has been claimed[5,7].

The amount of hard ($\tau_{hard}$) and soft ($\tau_{soft}$) lag between low ($E_l$) and high ($E_h$) photon energy can be approximated as the difference of $t_{acc}(E)$ and $t_{cool}(E)$ at the two energies[2], respectively,

$\tau_{hard} \approx t_{acc}(E_h) - t_{acc}(E_l)$
$= 9.65 \times 10^{-2} (1+z)^{3/2} \xi B^{-3/2} \delta^{-3/2} (E_h^{1/2} - E_l^{1/2})$ s  (3)

$\tau_{soft} \approx t_{cool}(E_l) - t_{cool}(E_h)$
$= 3.04 \times 10^3 (1+z)^{1/2} B^{-3/2} \delta^{-1/2} (E_l^{-1/2} - E_h^{-1/2})$ s  (4)

If the lag is measured with respective to the same energy, $\tau_{hard}(E)$ and $\tau_{soft}(E)$ become a function of photon energy. By fitting the observed energy-dependent lags with Equation (3) or (4), physical parameters of the emitting region can be constrained.

We apply equation (4) to fit the observed energy-dependent soft lags shown in Figure 3. The best fit gives $B\delta^{1/3} \sim 1.78$ Gauss. Zhang obtained $B\delta\xi^{-2/3} \sim 1.08 \times 10^{-3}$ Gauss by fitting the observed energy-dependent hard lags with equation (3) for an X-ray flare of Mrk~421 obtained with BeppoSAX[5]. Similar energy dependence of soft or hard lags were already obtained in Mrk~421 and other two similar TeV source PKS 2155-304 and Mrk 501[2,3,4,7,16,17,18]. Recently, Sato et al. showed the energy-dependent hard lags in another TeV source 1ES 1218+304 with SUZAKU data[19].

*This research is based on observations obtained with XMM-Newton, an ESA science mission with instruments and contributions directly funded by ESA Member States and NASA.*

1 Kirk J G, Rieger F, Mastichiadis A. Particle acceleration and synchrotron emission in blazar jets. Astron Astrophys, 1998, 333: 452－458
2 Zhang Y H, Treves A, Celotti A, et al. Four years of monitoring blazar PKS 2155-304 with BeppoSAX: probing the dynamics of the jet. Astrophys J, 2002, 572: 762－785
3 Zhang Y H, Treves A, Maraschi L, et al. XMM-Newton view of PKS 2155-304: hardness ratio and cross-correlation analysis of EPIC pn observations. Astrophys J, 2006, 637: 699－710
4 Kataoka J, Takahashi T, Makino F, et al. Variability pattern and the spectral evolution of the BL Lacertae Object PKS 2155-304. Astrophys J, 2000, 528: 243-253
5 Zhang Y H, Cross-spectral analysis of the X-ray variability of Markarian 421. Mon Not Astron Soc, 2002, 337: 609－618
6 Brinkmann W, Papadakis I E, den Herder J W A, et al. Temporal variability of Mrk 421 from XMM-Newton observations. Astron Astrophys, 2003, 402: 929-947
7 Ravasio M, Tagliaferri G, Ghisellini G, et al. Observing Mkn 421 with XMM-Newton: the EPIC-PN point of view. Astron Astrophys, 2004, 424:841-855
8 Albert J, Aliu E, Anderhub H, et al. Variable very high energy γ-ray emission from Markarian 501. Astrophys J, 2007, 669: 862-883
9 Edelson R A, Griffiths G, Markowitz A, et al. High temporal resolution XMM-Newton monitoring of PKS 2155-304. Astrophys J, 2001, 554: 274-280
10 Zhang Y H, Cagnoni I, Treves A, et al. The effects of periodically gapped time series on cross-correlation lag determinations. Astrophys J, 2004, 605: 98-104
11 Brinkmann W, Papadakis I E, Raeth C, et al. XMM-Newton timing mode observations of Mrk 421. Astron Astrophys, A&A, 2005, 443:397-411
12 Fossati G, Celotti A, Chiaberge M, et al. X-ray emission of Markarian 421: new clues from its spectral evolution. I. temporal analysis. Astrophys J, 2000, 541: 153-165
13 Zhang Y H, Treves A, Celotti A, et al. XMM-Newton view of PKS 2155-304: characterizing the X-ray variability properties with EPIC pn. Astrophys J, 2005, 629: 686-699
14 Massaro, E.; Perri, M.; Giommi, P.; et al. Log-parabolic spectra and particle acceleration in the BL Lac object Mkn 421: Spectral analysis of the complete BeppoSAX wide band X-ray data set. AA 2004, 413:489-503
15 Tramacere, A.; Massaro, F.; Cavaliere, A. Signatures of synchrotron emission and of electron acceleration in the X-ray spectra of Mrk 421 AA 2007, 466:521-529
16 Takahashi T, Tashiro M, Madejski G, et al. ASCA Observation of an X-Ray/TeV Flare from the BL Lacertae Object Markarian 421. Astrophys J, 1996, 470: L89-L92
17 Tanihata C, Urry C M, Takahashi T, et al. Variability timescales of TeV blazars observed in the ASCA continuous long-look X-ray monitoring. Astrophys J, 2001, 563: 569-581
18 Zhang Y H, Celotti A, Treves A, et al. Rapid X-ray variability of the BL Lacertae object PKS 2155-304. Astrophys J, 1999, 527: 719-732
19 Sato R, Kataoka J, Takahashi T, et al. Suzaku observation of the TeV blazar 1ES 1218+304: clues on particle acceleration in an extreme TeV Blazar. Astrophys J, 2008, 680: L9-L12